\documentclass[conference,pdfdvimx,a4paper]{IEEEtran}

\usepackage{cite,graphicx,color,url}
\usepackage{algorithm,algpseudocode}
\usepackage{amssymb,amsmath,mathtools}
\usepackage{amsfonts}

\usepackage{bm,bbm}
\usepackage{booktabs}
\usepackage[caption=false,font=footnotesize]{subfig}

\title{
	Packet-Loss-Tolerant Split Inference\\ for Delay-Sensitive Deep Learning \\in Lossy Wireless Networks}
\author{
	\IEEEauthorblockN{
		Sohei~Itahara\IEEEauthorrefmark{1},
		Takayuki~Nishio\IEEEauthorrefmark{2}, and
		Koji Yamamoto\IEEEauthorrefmark{1}}
	\IEEEauthorblockA{
        \IEEEauthorrefmark{1}Graduate School of Informatics, Kyoto University, Yoshida-honmachi, Sakyo-ku, Kyoto 606-8501 Japan\\
        \IEEEauthorrefmark{2}School of Engineering, Tokyo Institute of Technology, Ookayama, Meguro-ku, Tokyo, 152-8550, Japan\\
	}
    \IEEEauthorblockA{\IEEEauthorrefmark{2}nishio@ict.e.titech.ac.jp}
}

\begin{document}
\maketitle

\begin{abstract}
	The distributed inference framework is an emerging technology for real-time applications empowered by cutting-edge deep machine learning (ML) on resource-constrained Internet of things (IoT) devices.
	In distributed inference, computational tasks are offloaded from the IoT device to other devices or the edge server via lossy IoT networks.
	However, narrow-band and lossy IoT networks cause non-negligible packet losses and retransmissions, resulting in non-negligible communication latency.
	This study solves the problem of the incremental retransmission latency caused by packet loss in a lossy IoT network.
	We propose a split inference with no retransmissions (SI-NR) method that achieves high accuracy without any retransmissions, even when packet loss occurs.
	In SI-NR, the key idea is to train the ML model by emulating the packet loss by a dropout method, 
	which randomly drops the output of hidden units in a DNN layer.
	This enables the SI-NR system to obtain robustness against packet losses.
	Our ML experimental evaluation reveals that SI-NR obtains accurate predictions without packet retransmission at a packet loss rate of 60\%.
\end{abstract}

\section{Introduction}
\label{sec:introduction}
The Internet of things (IoT) is envisioned to provide many novel applications by combining the physical sensing and actuation of IoT devices with deep learning-based data analysis. 
Although deep learning technology is developing rapidly, it remains challenging to implement deep learning applications on resource-constrained IoT systems while satisfying the latency and privacy demands of IoT applications. 
For example, factory automation and smart grids require latency of less than 10 ms and 20 ms, respectively~\cite{philipp2017latency}.
On the other hand, in smart home applications, IoT sensors such as visual and audio sensors obtain privacy-sensitive data that should not be exposed~\cite{huichen2016iot}.

Distributed inference frameworks have been studied to address latency and privacy challenges of deep learning deployment in IoT. 
In the distributed inference framework for deep neural networks (DNNs), computationally expensive tasks are offloaded from the IoT devices to the locally located sink nodes or edge servers to reduce computation latency, communication latency, and the risk of data leakage compared with cloud computing.
In distributed DNN inference, a DNN is divided horizontally or vertically. 
The IoT devices and the edge server collaboratively process the portion of DNN, the so-called sub-DNN, by exchanging messages (e.g., outputs of sub-DNN) via IoT networks.
However, narrow-band and lossy IoT networks cause non-negligible packet losses, which can degrade the reliability of the inference and communication efficiency.

A straightforward approach to solve the packet loss problem is to employ a reliable retransmission protocol, such as the transmission control protocol (TCP).
However, retransmission causes non-negligible end-to-end communication latency, especially in narrow-band and lossy wireless networks. 
This latency problem motivated us to design a packet-loss-tolerant distributed inference method. 
Specifically, we aim to build an inference method in which a device and an edge server
collaboratively provide accurate predictions via lossy but
real-time connections, such as user datagram protocol (UDP).

\begin{figure}[!t]
	\centering
	\includegraphics[width = 0.35\textwidth]{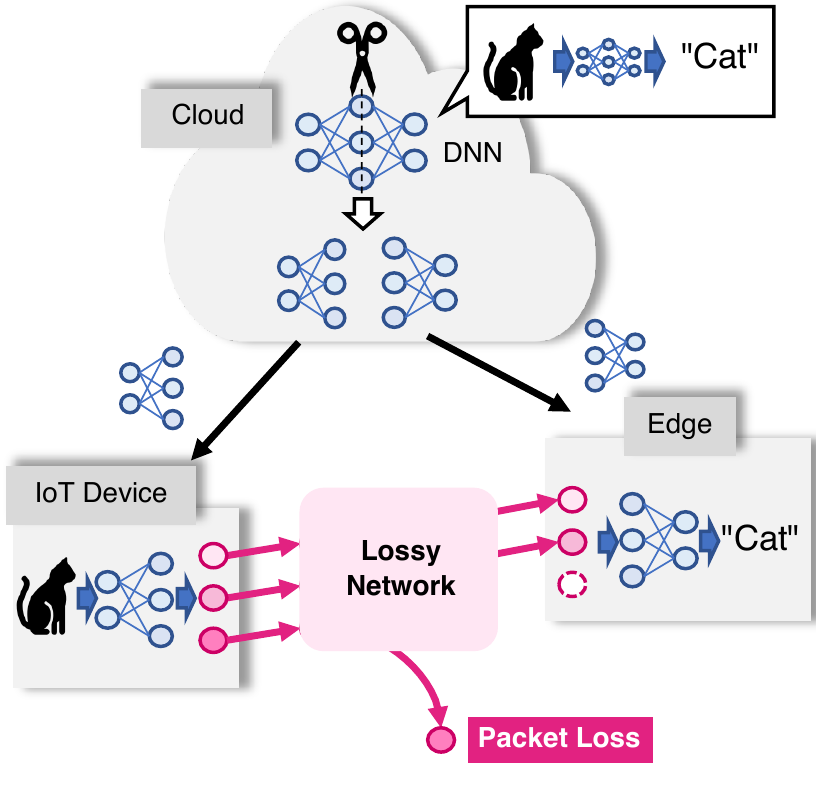}
	\caption{\textbf{Overview of the proposed split inference with no retransmissions.}
		Red arrows indicate the upload of intermediate representations.
		During the upload, intermediate representations are partially lost,  but the edge server does not require any retransmissions and processes forwarding using only the successfully transmitted representations.
	}
	\label{fig:abst}
\end{figure}

To this end, we propose a novel distributed inference framework called split inference with no retransmissions (SI-NR).
Fig.~\ref{fig:abst} shows the overview of SI-NR.
In SI-NR, even when the messages exchanged among the nodes are partially dropped owing to packet loss in the networks,
 one can obtain accurate inference results using only successfully received parts.
To obtain such packet-loss tolerance, our key idea is to train a DNN with emulating the packet loss by a dropout method, which randomly drops the output of hidden units in a DNN layer.
This training enables the model to predicts output accurately from dropped information.
We evaluate our method using an object classification task, CIFAR-10.
The results showed that our method achieved satisfactory accurate results even with a high packet loss rate.

\section{Related Work: Distributed Inference for Deep Neural Networks}
Distributed inference is a technique for performing an inference task collaboratively on multiple nodes.
This study focuses on distributed inference with DNNs, called distributed DNN inference~\cite{nicholas2015candeep,wang2018not,koda2020communication,liu2017appbooater,kim2017splitnet,zhao2018deeptings}.
In distributed inference frameworks, the DNN is horizontally~\cite{nicholas2015candeep,wang2018not,koda2020communication,liu2017appbooater} or vertically~\cite{kim2017splitnet,zhao2018deeptings} partitioned,
 and the IoT devices and the edge server collaboratively process DNN inference.
In the inference with a vertical model split, IoT devices and edge servers process sub-DNNs in parallel with shared input data to accelerate the computation by parallel processing.
On the other hand, in the inference with a horizontal model split, the IoT device and the edge server process their sub-DNNs sequentially.
First, an IoT device that obtains sensor data processes a sub-DNN.
Subsequently, other devices process their sub-DNNs with the outputs of other sub-DNNs in order of the original DNN.
Therefore, the inference with a horizontal model split may require longer computational time than the vertical model split.
However, the horizontal split involves less risk of data leakage than the vertical split because the entire raw data are processed by the source IoT device and not shared, whereas portions of raw data are shared in the vertical split.
Thus, this study focuses on the horizontal split to satisfy the data-privacy demands of IoT applications.

Details of the distributed DNN inference with a horizontal model split are explained as follows.
A well-trained DNN is divided into sub-DNNs by layers.
The input-side sub-DNN is stored by the IoT device, and the output-side sub-DNN is stored by the edge server.
The device obtains the output of the sub-DNN (i.e., the intermediate representations of the original DNN) from the raw input.
Next, the intermediate representation is transmitted to the edge server, and the server generates an inference result from the intermediate representation with its sub-DNN.
Some techniques~\cite{koda2020communication,bhardwaj2019memory} were proposed to reduce the communication cost of transmitting intermediate representations.
These methods reduced the amount of traffic in distributed inference, but the problem of packet loss in lossy IoT networks has not been addressed.
Unlike these works, this study aims to provide a packet loss-tolerant distributed inference method, which is an orthogonal work.

The impact of packet loss on DNN inference was studied in~\cite{liu2020improve}.
In~\cite{liu2020improve}, it is assumed that all DNN inference tasks are offloaded to the server, and the input images are corrupted owing to UDP packet loss.
This study revealed that, as reported for conventional centralized inference~\cite{dodge2016understanding}, the DNN can provide accurate predictions for partially corrupted images in environments where the packet loss rate is less than 1\%.
This work demonstrated the feasibility and effectiveness of employing UDP for AI-empowered time-critical applications in non-lossy networks and has motivated us to build a distributed inference method that can withstand harsh environments, i.e., lossy IoT networks where the packet loss rate could be more than several tens of percent.

\section{Proposed Method: Split Inference With No Retransmissions}
\label{sec:proposed_method}

\begin{figure}[!t]
	\centering
	\includegraphics[width = 0.35\textwidth]{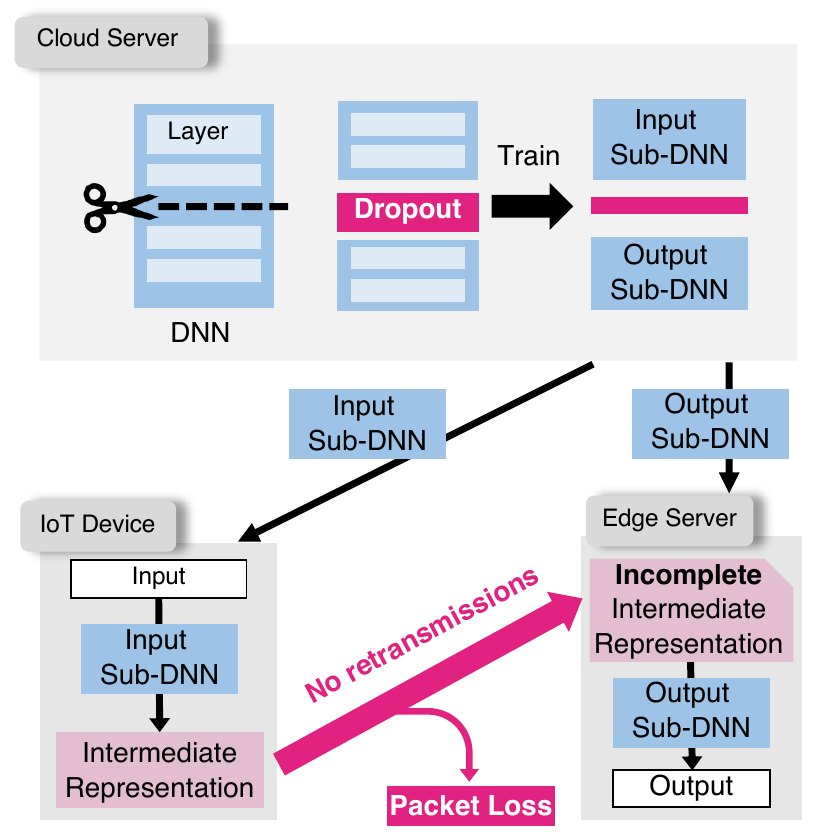}
	\caption{\textbf{Detailed procedure of the proposed split inference with no retransmissions.}
		Red arrows indicate the upload of the intermediate representations from the device to the edge server via the UDP connection, which does not retransmit dropped packets.
		The edge server obtains prediction results using only the successfully transmitted intermediate representations.
	}
	\label{fig:proposed_method}
\end{figure}
As shown in Fig.~\ref{fig:proposed_method}, the system model consists of a cloud server, an edge server, and an IoT device.
The cloud server trains a DNN model to solve the inference tasks generated on the IoT device.
The portions of the well-trained DNN (sub-DNNs) are distributed to the IoT device and the edge server.
The IoT device and edge server solve inference tasks collaboratively with distributed sub-DNNs.
The device has a much weaker computational capacity than the cloud server and the edge server.
We assume that the cloud server holds a sufficient amount of data and computational capacity to train the DNN model.

The proposed distributed inference method called split inference with no retransmissions (SI-NR) consists of two phases: a model training phase and a split inference (SI) phase.
In the model training phase, the cloud server trains a DNN using a preobtained dataset.
To achieve robustness to packet loss in the SI phase, the DNN is trained by emulating the packet loss using a dropout method.
Details of the model training are described in the next section. 
The trained DNN is horizontally divided into an input-side and output-side of the original DNN, which are called input sub-DNN and output sub-DNN, as shown in Fig.~\ref{fig:proposed_method}.
The input and output sub-DNNs are sent to the device and the edge server, respectively.
The SI phase is conducted when an inference task is generated in the IoT device.
In the SI phase, the device generates intermediate representations using the input sub-DNN and sends the intermediate representations to the edge server via the UDP connection, which does not retransmit dropped packets.
The edge server obtains the prediction results with the output sub-DNN by inputting the successfully received intermediate representations from the IoT device. 

This study assumes that the transmitted intermediate representations are probabilistically dropped in the communication channel owing to packet loss. 
More formally, considering that the device sends a vector $\bm{x}$ via a communication channel with a packet loss rate $p$,
the edge server successfully receives a vector $f^\mathrm{c}(\bm{x},p)$ denoted as follows:
\begin{align}
	\label{equ:communication_channel}
	f^\mathrm{c}(\bm{x},p) = \bm{x} \odot \bm{m}(p),
\end{align}
where operator $\odot$ indicates the element-wise product and $\bm{m}(p)$ is a binary vector derived from the Bernoulli distribution with an expected value of $1-p$.

In a real-world communication system, the vector of the intermediate representations $\bm{x}$ is divided into multiple packets and then transmitted. 
Therefore, when a packet is dropped, the consecutive elements of $\bm{x}$ are lost. 
To avoid such burst loss, the device shuffles the elements of the vector and stores them in packets.
The edge server constructs vector $\bm{x}$ from successfully received packets, which results in~\eqref{equ:communication_channel}.

\subsection{Training a packet-loss-tolerant model}
We leverage a dropout technique~\cite{hinton2012improving} to make a DNN model tolerant to packet loss.
The dropout technique was originally proposed as an efficient regularization method for DNNs, and it randomly drops the output of hidden units in a DNN layer.
The dropout can train the DNN for longer periods without overfitting, which improves the test accuracy~\cite{hinton2012improving}. 
Thus, the dropout technique has been used in various DNN architectures. 
In this study, we leverage the dropout to emulate packet loss in the model training phase and enable the model to perform accurate inferences from incomplete intermediate representations. 

In each training iteration with the dropout technique, the outputs of the hidden units are set to zero using a dropout layer with a dropout rate $r$. 
In addition to omitting the hidden unit outputs, the surviving (not omitted) hidden units are multiplied by $1/(1-r)$.
More formally, the dropout behavior $f^\mathrm{d}(\cdot,r)$ is represented as follows:
\begin{align}
	\label{equ:dropout}
	\bm{x}_{i+1} = f^\mathrm{d}(\bm{y}_i,r) =
	\frac{1}{1-r}\bm{y}_i \odot \bm{m}(r),
\end{align}
where $\bm{y}_i$ is the hidden unit of the $i$th layer, and $\bm{x}_{i+1}$ is the input of the $i+1$th layer.
Comparing equations~\eqref{equ:communication_channel} and~\eqref{equ:dropout}, we find that the drops of intermediate representations due to packet loss can be emulated by the dropout technique in the training phase.
Therefore, the model trained with the dropout technique can provide accurate inferences even when the intermediate representations are dropped because of the packet loss.
The detailed training procedure is described as follows.

The cloud server first determines a division point, which separates a DNN into input and output sub-DNNs.
The division point is determined by considering the computational capacity of the IoT device and the edge server.
When the DNN is separated close to the input layer, the input sub-DNN becomes small, which requires less computational power on the IoT device because more computational tasks are offloaded to the edge server.
A dropout layer is inserted into the division point, and then the model with the dropout layer is trained with the preobtained datasets.
Finally, the well-trained DNN is horizontally divided at the division point into the input and output sub-DNNs by removing the dropout layer.
More formally, the entire DNN model is denoted as $f^\mathrm{ent}(\cdot\vert \bm{w}^\mathrm{ent})$, where $\bm{w}^\mathrm{ent}$ are trainable parameters.
The entire DNN consists of two DNNs connected by the dropout layer as follows:
\begin{align}
	f^\mathrm{ent}(\cdot \vert \bm{w}^\mathrm{ent}) =
	f^\mathrm{out}(\cdot\vert \bm{w}^\mathrm{out})\circ f^\mathrm{dr}(\cdot,r) \circ f^\mathrm{in}(\cdot \vert \bm{w}^\mathrm{in}),
\end{align}
where $f^\mathrm{in}(\cdot\vert \bm{w}^\mathrm{in})$, $f^\mathrm{out}(\cdot \vert \bm{w}^\mathrm{out})$, and $f^\mathrm{dr}(\cdot,r)$ indicate the input sub-DNN, output sub-DNN, and dropout layer with a dropout rate of $r$, respectively.
The input sub-DNN $f^\mathrm{in}(\cdot\vert \bm{w}^\mathrm{in})$ and an output sub-DNN $f^\mathrm{out}(\cdot\vert \bm{w}^\mathrm{out})$ are distributed to the IoT device and the edge server, respectively.

Training with a larger dropout rate implies that the DNN is trained to adapt to a more lossy communication channel, which improves the packet loss tolerance.
On the other hand, as mentioned in~\cite{hinton2012improving}, a larger training dropout rate degrades the achievable model performance, 
i.e., performance without any packet loss. 
Therefore, the dropout rate is selected considering both the packet loss rate of the channel and the model performance requirements.

\subsection{Details of the split inference phase}
The SI phase is conducted when an inference task with input $\bm{x}$ is generated in the IoT device.
First, the device generates the intermediate representations $\bm{y}^\mathrm{int}$ as follows:
\begin{align}
	\bm{y}^\mathrm{int} \coloneqq \{y^\mathrm{int}_i \mid y^\mathrm{int}_i \in f^\mathrm{in}(\bm{x} \mid \bm{w}^\mathrm{in})\}.
\end{align}
The elements of vector $\bm{y}^\mathrm{int}$ are stored in packets to transmit them to the edge server.
As mentioned in Section~\ref{sec:proposed_method}, if the elements are stored in packets in order, a packet loss causes the loss of consecutive elements, which is not consistent with the assumption of packet loss in the model training.
Instead, the device permutates the elements randomly and stores them in packets. 
A packet $\bm{p}_i$ is represented as follows:
\begin{align}
	\bm{p}_i \coloneqq \{y^\mathrm{int}_{k_j} \mid i \leq  j < i+s\},
\end{align}
 where $k_j$ and $s$ are the permuted identification of the element and the number of elements stored in a packet, respectively. 
The edge server reconstructs the vector of intermediate representations from a subset of transmitted packets $\bm{P}^\mathrm{r}$, where
\begin{align}
	\bm{P}^\mathrm{r} = \{\bm{p}_i \mid \bm{p}_i\, \text{is received successfully}\}. 
\end{align}
Next, the edge server multiplies $1/(1-p)$ to the vector, where $p$ is the packet loss ratio, similarly to the dropout layer in the training phase.
Thus, the reconstructed vector is calculated as
\begin{align}
	\bm{y}^{\mathrm{int}'} = \frac{1}{1-p}\bm{y}^\mathrm{int} \odot \bm{m}(p).
\end{align}
Let $f^{\mathrm{c}'}(\cdot,p)$ denote the communication channel with a packet loss rate $p$ and the multiplication of $1/(1-p)$. 
The prediction result $\bm{y}^\mathrm{pred}$ can be written as
\begin{align}
	\bm{y}^\mathrm{pred} =
	f^\mathrm{out}(\cdot\vert \bm{w}^\mathrm{out})\circ f^{\mathrm{c}'}(\cdot,p) \circ f^\mathrm{in}(\bm{x} \vert \bm{w}^\mathrm{in}).
\end{align}
If $f^\mathrm{dr} (\cdot,r)$ in the training phase is close to $f^{\mathrm{c}'}(\cdot,p)$ in the SI phase, the model is expected to  accurately predict from the incomplete intermediate representations.

\subsection{Mathematical analysis of latency and accuracy}
This section provides a mathematical analysis of the transmission latency and prediction accuracy of two methods: the proposed method and the split inference with packet retransmission.
Specifically, it is shown that the proposed split inference can guarantee latency, whereas the split inference with packet retransmission cannot guarantee latency but achieves a higher prediction accuracy than the proposed method.

In the proposed method, if $n^\mathrm{t}$ packets are transmitted using a communication channel with a packet loss rate $p$,
the probability density function (PDF) of the number of received packets is calculated as follows:
\begin{align}
	f^\mathrm{rec}(n) =
	\begin{cases}
		{n^\mathrm{t} \choose n} p^{n^\mathrm{t}-n}(1-p)^n, & \text{if}\ 0 \leq n \leq n^\mathrm{t}; \\
		0,                                                  & \text{otherwise}.
	\end{cases}
\end{align}
The expected number of received packets is denoted by $(1-p)n^\mathrm{t}$.
Let $n^\mathrm{int}$ and $\text{acc}(\alpha)$ denote the number of packets that contain intermediate representations and the accuracy using $\alpha\%$ of intermediate representations, respectively.
Then, we can obtain the PDF of the accuracy as
\begin{align}
	\label{equ:prop_acc}
	f^\mathrm{acc}(\mathrm{acc}(\alpha)) = f^\mathrm{rec}\left(\frac{\alpha}{100} n^\mathrm{int}\right).
\end{align}
Assuming throughput $b$ and packet size $l$, the PDF of latency is calculated as
\begin{align}
	\label{equ:prop_delay}
	f^\mathrm{latency}(t) = \delta(t-n^\mathrm{t}T),
\end{align}
where $T = l/b$.
The variance of latency is $0$, and 
the expected value of latency is denoted as $n^\mathrm{t}T$.

In contrast, all the transmitted packets are received when using packet retransmission; therefore, the PDF of the accuracy is 
\begin{align}
	\label{equ:prev_acc}
	f^\mathrm{acc}(\mathrm{acc}(\alpha)) = \delta(\alpha - 100).
\end{align}
PDF of latency is
\begin{align}
	\label{equ:prev_delay}
	f^\mathrm{latency}(nT) =
	\begin{cases}
		{n-1 \choose n^\mathrm{t}-1} p^{n-n^\mathrm{t}}(1-p)^{n^\mathrm{t}}, & \text{if}\ n \geq  n^\mathrm{t}; \\
		0,                                                                   & \text{otherwise}.
	\end{cases}
\end{align}

Comparing the PDFs of communication latency in two methods, the proposed split inference method \eqref{equ:prop_delay} and split inference with packet retransmission \eqref{equ:prev_delay}, 
 the variance of the proposed method was 0, whereas that of the split inference with packet retransmission was larger than 0.
In other words, the proposed method can guarantee latency, whereas split inference with packet retransmission cannot guarantee latency.
On the other hand, comparing the PDFs of the accuracy, \eqref{equ:prop_acc} and \eqref{equ:prev_acc}, the variance of accuracy of the proposed method was larger than 0, whereas that of the split inference with packet retransmission was 0.
However, the following machine learning (ML) evaluation reveals that the range of the accuracy of the proposed method was less than 1\% under a packet loss rate of several tens of percent.

\section{Evaluation}
\label{sec:evaluation}
\subsection{Setup}
\label{ssec:setup}
\begin{figure}[!t]
	\centering
	\includegraphics[width=0.28\textwidth]{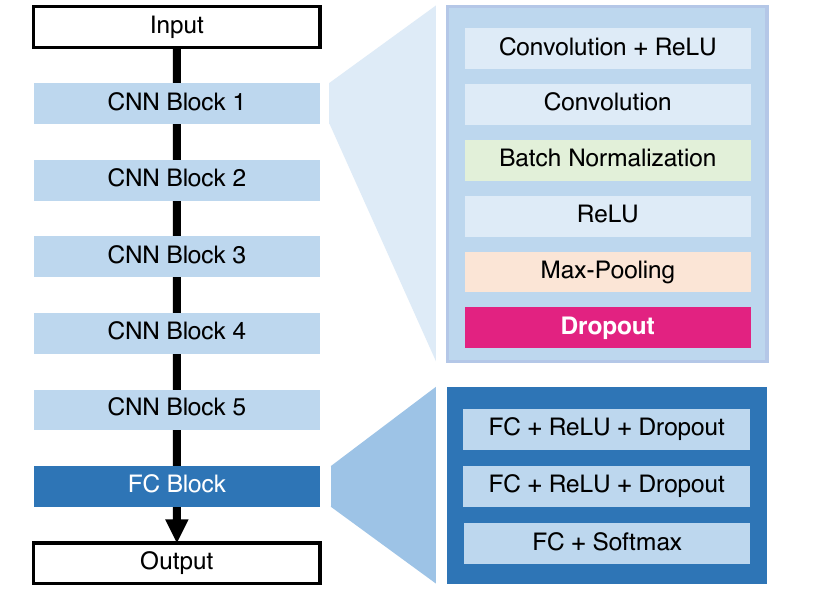}
	\caption{Architecture of DNN.
		Each CNN block consists of two or three convolutional, batch normalization, max-pooling, and dropout layers.
		The FC block consists of three FC layers.}
	\label{fig:ML_model}
\end{figure}

\begin{table}[!t]
	\caption{Detailed architecture of DNN.}
	\centering
	\scalebox{0.9}{
	\begin{tabular}{ccccccc}
		\toprule
		Block & Num.         & Num.            & Data size of intermediate \\
		ID    & conv. layers & output channels & representation           \\
		\midrule
		1     & 2            & 64              & 65.5\,kB                 \\
		2     & 2            & 128             & 32.8\,kB                 \\
		3     & 3            & 256             & 16.4\,kB                 \\
		4     & 3            & 512             & 8.2\,kB                  \\
		5     & 3            & 512             & 2.0\,kB                  \\
		\bottomrule
	\end{tabular}
	}
	\label{tab:CIFAR_CNN}
\end{table}

We conducted ML experiments as follows.
An IoT device and an edge server were assumed to be connected via a wireless channel, where packets were randomly dropped with probability $p$.
Hence, the elements of the intermediate representation vector transmitted by the IoT device were randomly dropped.
To calculate the communication latency, the packet size and throughput of the wireless channel (including MAC and network layer overheads) were set to 500\,bytes and 9.0\,Mbit/s.

We used an image recognition dataset, CIFAR-10\footnote{\url{https://www.cs.toronto.edu/~kriz/cifar.html}}, with 50,000 training and 10,000 testing images that represented 10 image classes, such as ``dog'' and ``ship.''
In the model training phase, the cloud server used all training dataset.
The test dataset was used to evaluate the inference performance of the SI phase of the proposed method.

The DNN model used in the experiments is shown in Fig.~\ref{fig:ML_model}.
The model was designed with reference to VGG16~\cite{simonyan2014very}, which consisted of five convolutional blocks and a fully connected (FC) block.
Each convolutional block included two or three $3 \times 3$ convolution layers activated by the rectified linear unit (ReLU), and the block was followed by $2 \times 2$ max-pooling and dropout layers.
In each convolutional block, the convolutional layers had the same number of output channels.
Additionally, one of the two convolutional layers was followed by the batch normalization layer.
Table~\ref{tab:CIFAR_CNN} describes each CNN block, such as its number of convolutional layers and the data size of intermediate representations.
The FC block consists of three FC layers (256 and 128 units with ReLU activation and dropout, and 10 other units activated by softmax).

The training dataset was divided into updating and validation datasets in a ratio of 9:1.
The DNN model was updated using only the updating dataset for multiple epochs.
In each epoch, the model is evaluated using the validation dataset.
The training is completed if there are 150 epochs performed, or if the validation loss is increased after 20 epochs consecutively, which indicates that the model is starting to overfit.
The same dropout rate was used to train all dropout layers.
The Adam optimizer, the training rate of 0.001, and the mini-batch size of 128 were selected as hyperparameters.

To evaluate the accuracy of the split inference, the CNN was divided at a specific CNN block.
Considering that the CNN is divided at the CNN block $i$, the input sub-DNN consists of the CNN blocks $1,2,\dots, i$, and the output sub-DNN consists of the CNN blocks $i+1, i+2, \dots, 5$, and the FC block.
The packet loss in the communication channel is emulated by the dropout.
In particular, the dropout rate of the CNN block $i$ is set to the packet loss rate $p$, which ranges from 0 to 0.9.
The dropout rates of the other CNN blocks and the FC block were set to 0.

\subsection{Results}
\subsubsection{Cumulative distribution function of the accuracy and latency}
\begin{figure}[!t]
	\centering
	\subfloat[Accuracy]{\includegraphics[width=0.22\textwidth]{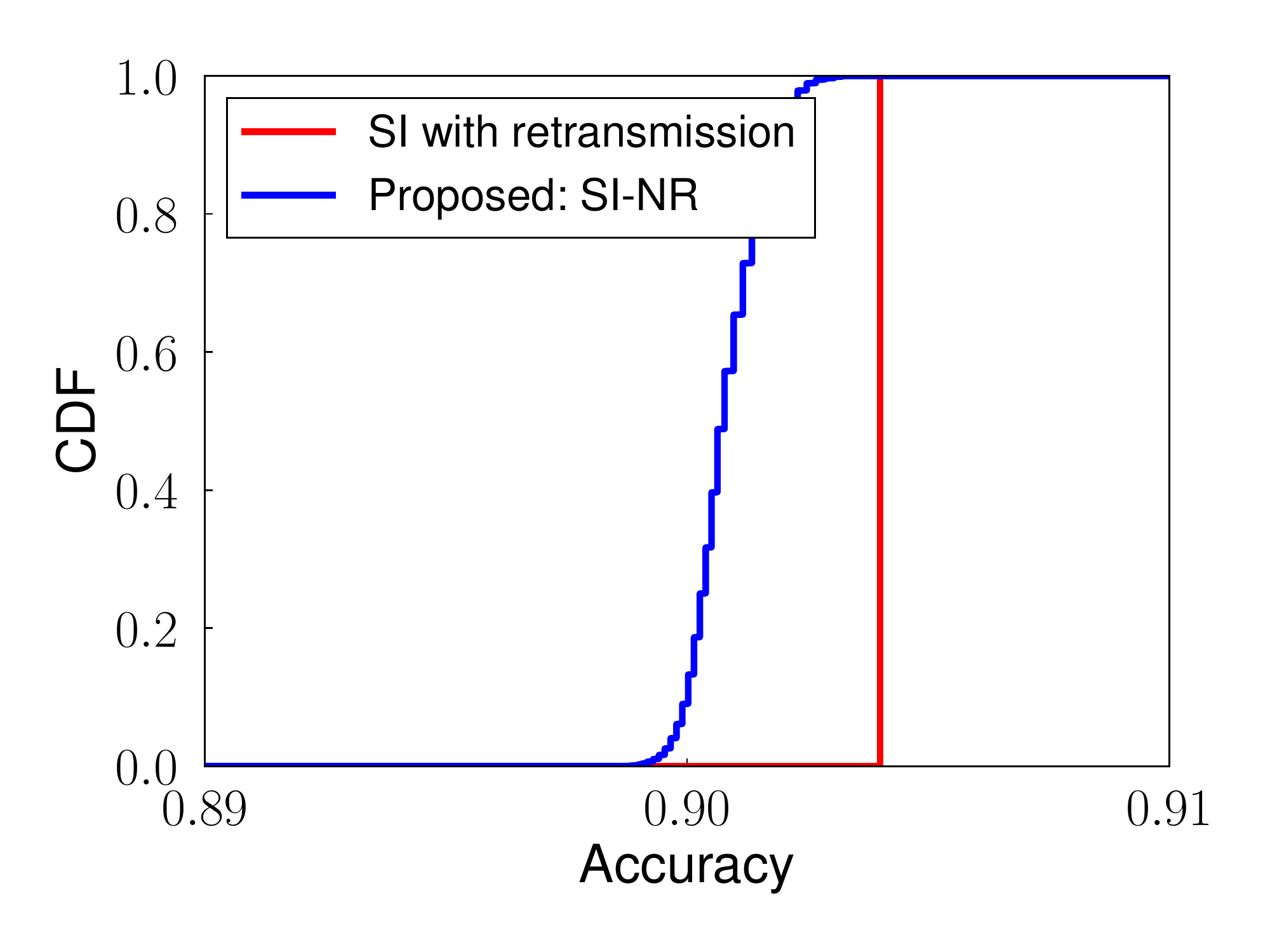}}
	\subfloat[Communication latency]{\includegraphics[width=0.22\textwidth]{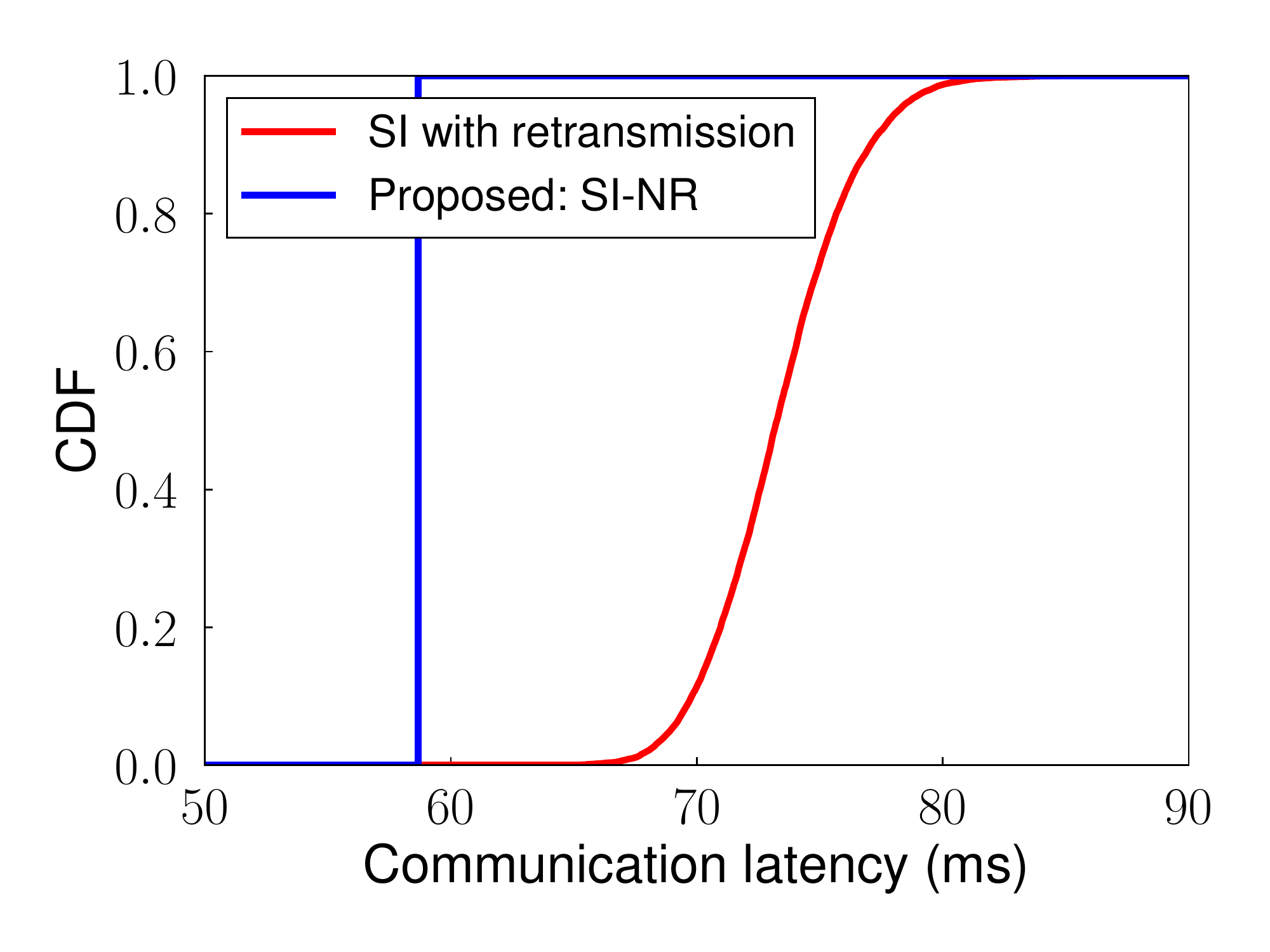}}
	\caption{Cumulative distribution function of the accuracy and communication latency for the conventional inference with packet retransmission and the proposed split inference with no retransmissions.
	The number of packets to be transmitted is the same for the two methods.}
	\label{fig:prob_same_rx}
\end{figure}

Fig.~\ref{fig:prob_same_rx} illustrates the cumulative distribution function (CDF) of the accuracy and latency for the proposed SI and conventional SI with retransmission, respectively.
The CDF is obtained from \eqref{equ:prop_acc}, \eqref{equ:prop_delay}, \eqref{equ:prev_acc}, and \eqref{equ:prev_delay}, with the parameters described in Section~\ref{ssec:setup} and the packet loss rate of 0.2. 
The model was divided at the CNN block 1.
The number of transmitted packets was the same in both methods, but the proposed method received fewer packets owing to the non-retransmission policy.

In Fig.~\ref{fig:prob_same_rx}~(a), the accuracy of the SI-NR is  approximately 0.9, whereas that of the conventional SI is stable.
This is because the proposed method does not perform any retransmission, and thus a part of the intermediate representation is stochastically dropped owing to packet loss.
In contrast, the conventional method always obtains complete intermediate representations owing to the retransmission.
On the one hand, the packet loss-induced accuracy degradation in the proposed method is less than 1\%. 
This result indicates that the proposed method achieves packet loss tolerance, which is verified in more detail in the following sections.

On the other hand, Fig.~\ref{fig:prob_same_rx}~(b) shows that the communication latency of the proposed method is stable and smaller than that of the conventional method.
This is because in the proposed method, the device sends only a
predefined number of packets, whereas the conventional method transmits a larger number of packets until all packets are successfully received.
These results demonstrate that the proposed method achieves a guaranteed small communication latency with a slight degradation in model performance.

\subsubsection{Impact of the training dropout rate on the achievable model accuracy}
\begin{figure}[!t]
	\centering
	\includegraphics[width = 0.3\textwidth]{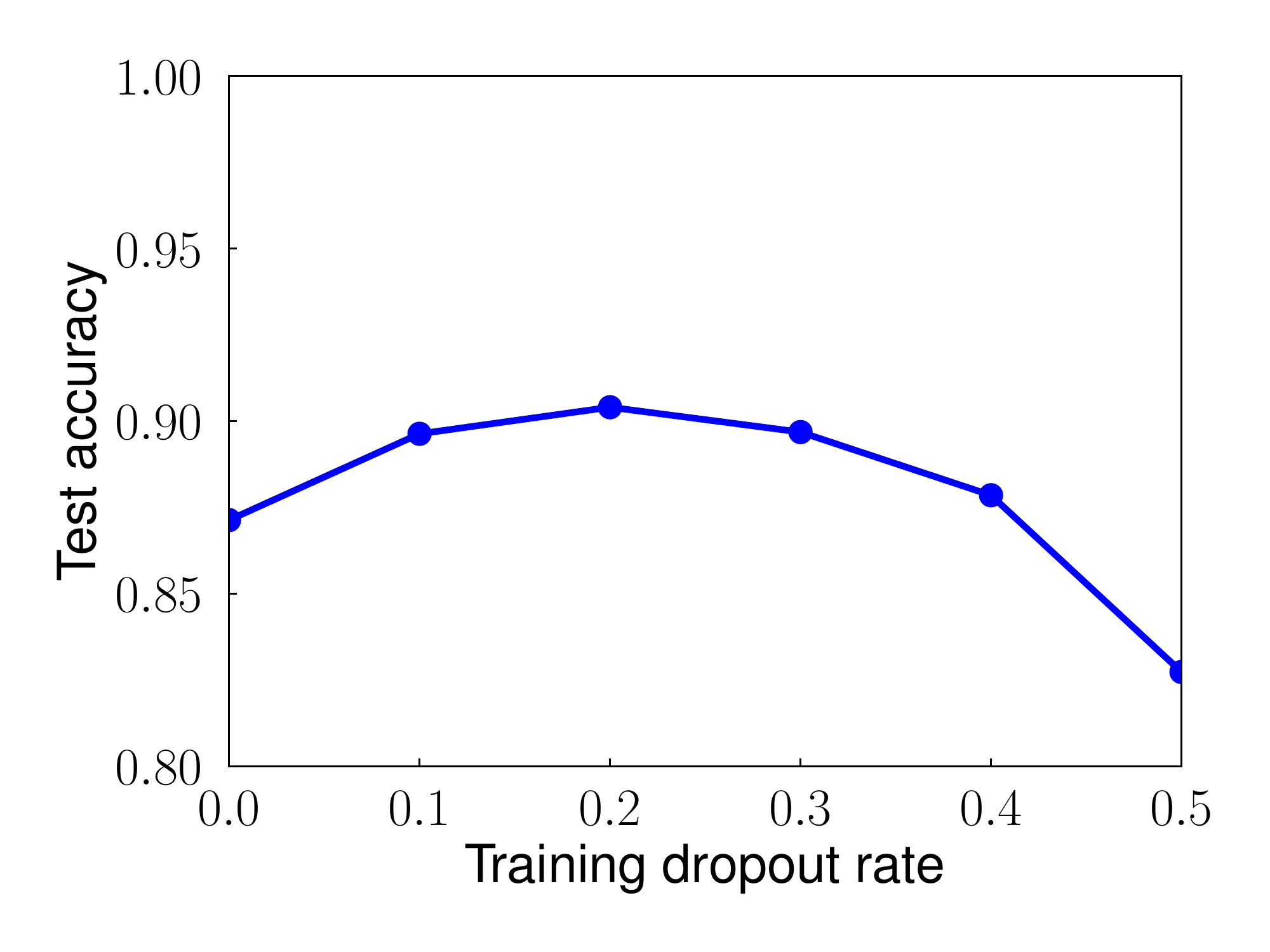}
	\caption{Test accuracy without any packet loss as a function of training dropout rate.
		The training dropout rate of 0.0 indicates the model trained without dropout.
	}
	\label{fig:cifar_train_dr_acc}
\end{figure}

Fig.~\ref{fig:cifar_train_dr_acc} shows the test accuracy without any packet loss.
Consistently with a previous study~\cite{hinton2012improving}, training with dropout improves testing accuracy, 
owing to its regularization capability. 
In the ML task we adopted in this evaluation, the model trained with a dropout rate of 0.2 achieved the highest test accuracy.
As mentioned in~\cite{hinton2012improving}, the dropout rate that achieves the highest performance depends on the ML task and DNN architecture.
Therefore, it should be tuned on the validation dataset in the training phase while considering the packet loss rate of the lossy networks.

\subsubsection{Impact of the training dropout rate on the prediction accuracy of split inference}
\begin{figure}[!t]
	\centering
	\includegraphics[width = 0.3\textwidth]{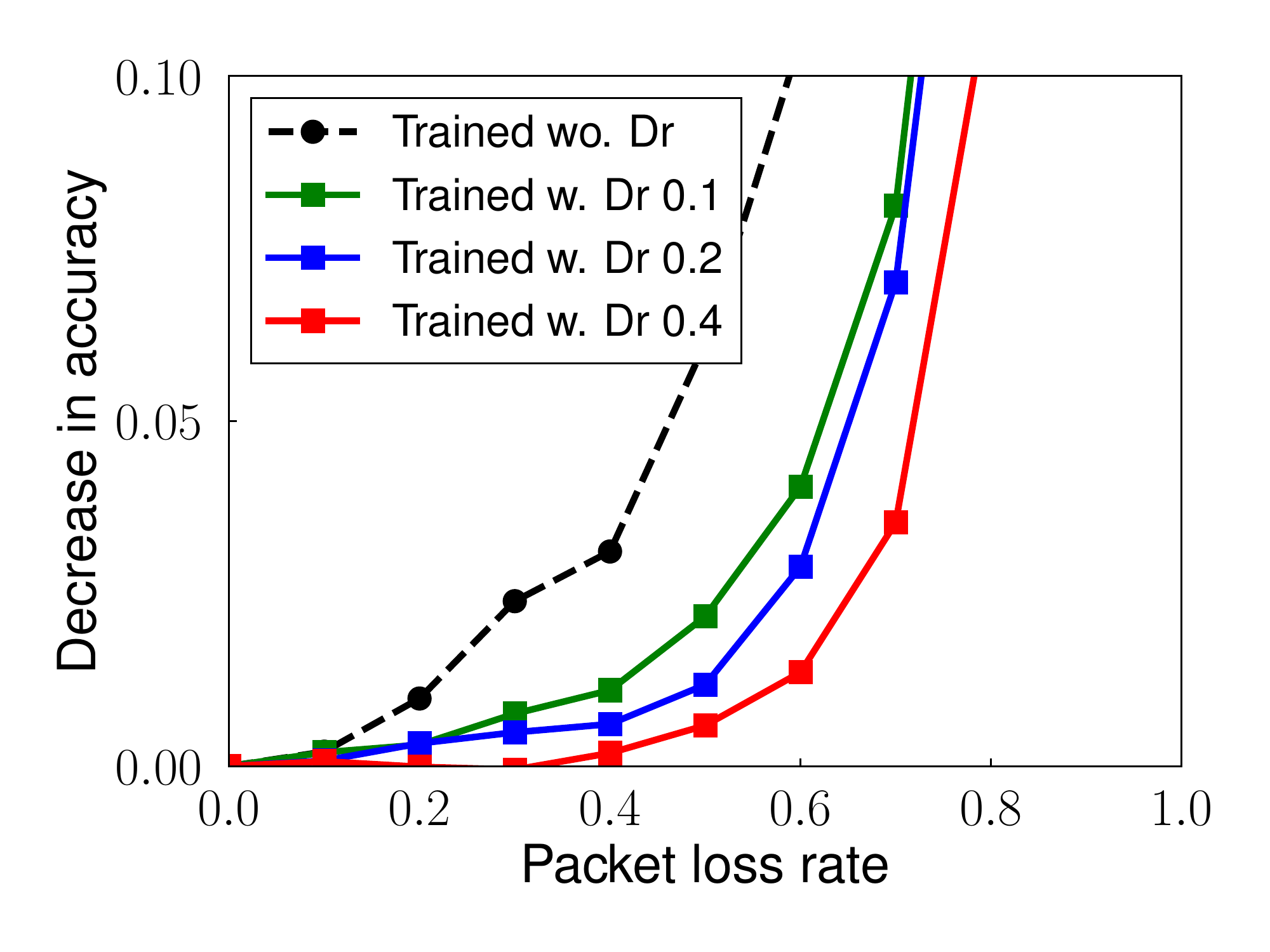}
	\caption{Effect of the training dropout rate on the model performance degradation.
		The dotted black line indicates the results obtained using the DNN trained without any dropout layer.
		The green, blue, and red lines indicate the results using the DNN trained with dropout rates of 0.1, 0.2, and 0.4, respectively.
	}
	\label{fig:cifar_inf_dr_comp_tr_dr}
\end{figure}

Fig.~\ref{fig:cifar_inf_dr_comp_tr_dr} shows the decreases in test accuracy in the method without any packet loss as a function of packet loss rate, when the DNN is divided at block 1.
As the packet loss rate increases, the model accuracy degrades.
The proposed method with a larger training dropout rate better mitigated the performance degradation.
In particular, the performance of the model trained without dropout was degraded by more than 0.1, when more than 60\% of the packets were dropped.
The model trained with a dropout rate of 0.4 showed only a slight performance degradation. 
This result demonstrates that model training with the dropout technique significantly improves the packet loss tolerance of the split model.

\subsubsection{Effect of the division point on the prediction accuracy}
\begin{figure}[!t]
	\centering
	\includegraphics[width = 0.3\textwidth]{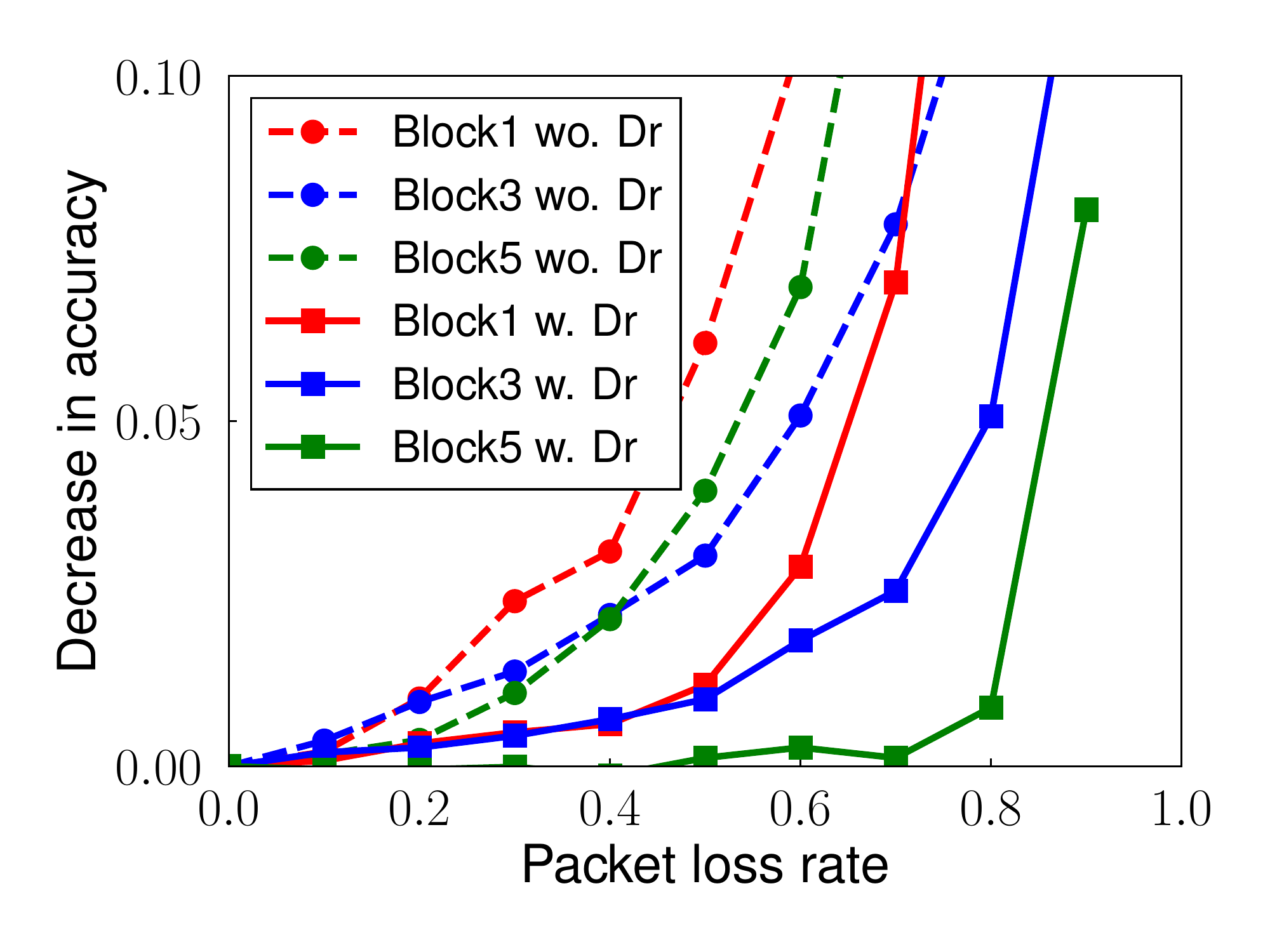}
	\caption{Effect of the division point on the model performance degradation.
		The dotted lines and solid lines indicate the results obtained using the model trained without any dropout layer and with a dropout rate of 0.2, respectively.
		Different colors indicate different division layers that split the DNN into input and output sub-DNNs.
	}
	\label{fig:cifar_inf_dr_acc}
\end{figure}

Fig.~\ref{fig:cifar_inf_dr_acc} shows the decrease in the test accuracy for the model divided at different points. 
Consistently with the results in the above section, the DNN trained with dropout achieved less accuracy degradation than that without dropout, regardless of the division points.
On the one hand, the DNN divided at the division point closer to the output layer achieved less performance degradation for a high packer loss rate. 
In addition to the increase in the robustness against packet loss, as the division point approaches the output layer of the original DNN, the data size of the intermediate representation became generally smaller, as shown in Table~\ref{tab:CIFAR_CNN}, which reduced  communication latency. 
However, it also increased the size of the input sub-DNN, which should be processed by the resource-constrained IoT device, thereby increasing the computation latency.
Therefore, the division point should be carefully determined to maximize the model accuracy under the constraints of the total latency.

\section{Conclusion}
\label{sec:conclusion}
We have presented a packet loss tolerant split inference method, SI-NR.
In SI-NR, the key idea is to train the ML model by emulating the packet loss by a dropout method so that the model gains packet loss tolerance.
Our experimental ML evaluation reveals that SI-NR obtains an accurate prediction without packet retransmission under a high packet loss rate.
An interesting area for future work is an optimization framework that determines the division point and the dropout rate to maximize the model accuracy in lossy wireless networks under the constraints of the total latency of communication and computation.

\section*{Acknowledgement}
This work was supported in part by JST PRESTO Grant Number JPMJPR2035.
\bibliographystyle{IEEEtran}
\bibliography{arXiv}

\end{document}